\documentclass[prb,preprint,longbibliography]{revtex4-1} 

\usepackage{amsmath} 

\usepackage{graphicx} 
\usepackage{epstopdf} 

\begin{document}

\title{Incorporating learning goals about modeling into an upper-division physics laboratory experiment}
\author{Benjamin M. Zwickl}
\altaffiliation[Current Address: ]{School of Physics and Astronomy, Rochester Institute of Technology, Rochester, NY 14623}
\affiliation{Department of Physics, University of Colorado Boulder, Boulder, CO 80309}
\email{benjamin.m.zwickl@rit.edu} 
\author{Noah Finkelstein}
\affiliation{Department of Physics, University of Colorado Boulder, Boulder, CO 80309}
\author{H. J. Lewandowski}
\altaffiliation[Also at ]{JILA, University of Colorado Boulder, Boulder, CO 80309} 
\affiliation{Department of Physics, University of Colorado Boulder, Boulder, CO 80309}
\date{\today}

\begin{abstract}
Implementing a laboratory activity involves a complex interplay among learning goals, available resources, feedback about the existing course, best practices for teaching, and an overall philosophy about teaching labs.  Building on our previous work, which described a process of transforming an entire lab course, we now turn our attention to how an individual lab activity on the polarization of light was redesigned to include a renewed emphasis on one broad learning goal: modeling.  By using this common optics lab as a concrete case study of a broadly applicable approach, we highlight many aspects of the activity development and show how modeling was used to integrate sophisticated conceptual and quantitative reasoning into the experimental process through the various aspects of modeling: constructing models, making predictions, interpreting data, comparing measurements with predictions, and refining models.  One significant outcome is a natural way to integrate an analysis and discussion of systematic error into a lab activity.
\end{abstract}

\maketitle

\section{Introduction}

Lab courses are complex environments that present instructors with many options, both regarding the overarching features of the course and the finer details specific to each lab activity.  Course-scale decisions include: \textit{What labs should students do?  What physics topics should they cover?  What equipment should be purchased?  How many days and hours should each lab take?  What kinds of collaboration and group work should be encouraged?  What work will be submitted for grading?}  When creating individual activities, many finer details arise: \textit{What kinds of prompts should be in the lab guide?  How much of the experiment should be set up in advance?  How much of the theory should students understand?}  While there are probably many answers to these questions that result in quality student learning, the variety of options can seem daunting to anyone transforming an existing laboratory course or building one from the ground up.\cite{Carlson1986}  

At the University of Colorado Boulder (CU), we have undertaken a systematic research-based approach to establish goals for laboratory instruction, transform our courses, and conduct research into teaching and learning experimental and research skills in laboratory courses.  In our previous article, \emph{The Process of Transforming an Advanced Lab Course}, we presented an overview of learning goals, curricula, and assessments that were developed and used in our advanced lab course.\cite{Zwickl2013}  Through discussions with over twenty CU physics faculty, we established four broad goals for our lab courses: modeling (of physical systems and measurement tools), design (of experiments and apparatus), communication, and technical lab skills (i.e., data analysis, measurement and automation, and the use of standard lab equipment).  While each of these broad goals is important and encompasses a range of skills and abilities, this article clarifies how we implemented a set of modeling-related goals in one particular laboratory activity.   Further, we show how an explicit discussion of modeling can integrate sophisticated conceptual and quantitative reasoning into the experimental process.  

We highlight many aspects of the development of activities for a common optics lab on the polarization of light.  The process serves as a case study of a broadly applicable approach. We begin by taking a big-picture look at a teaching philosophy for laboratory courses, then get more specific by discussing modeling in the context of a laboratory course, then detail some lab-specific learning goals for the polarization of light lab, and finally implement the learning goals in an activity.  We also provide a graphical schematic of the modeling process (Fig.\ \ref{fig:Modeling_framework}) that describes modeling as a series of interconnected steps relating the real world apparatus, model construction, predictions, comparison, and model refinement.  We have used Fig.\ \ref{fig:Modeling_framework} extensively in our own work to elucidate the relationship between theory and measurement in experimental physics.

\section{A teaching philosophy for labs}

We start by laying out a clear teaching philosophy for upper-division labs.  We start with this broad perspective because it helps us make  decisions about equipment and apparatus.  Polarized light is a topic that has been studied for over 200 years, so it is a fair question to ask if this lab is relevant today.  As we describe our philosophy in more detail, it will become clear why we think the polarization of light lab is still an excellent choice for our advanced lab course.  Our philosophy, summarized in Table\ \ref{tab:URE_vs_Lab_course}, is that a good lab is one that enables us to meet our consensus learning goals and prepares students to engage in research.  We are not treating our labs as a substitute for research, but as preparation for research.  This philosophy helps us answer big questions about the entire lab course: \textit{What labs should students do?  (Is the polarization of light lab important?)  What equipment should be purchased?  (Or is what we have good enough?)  How long should the lab take?}   

Because the lab course is a many-to-one student-to-instructor environment, the course should focus on content and scientific practices that are widely relevant to research and amenable to that instructional environment.  For instance, our faculty identified particular research abilities, such as computer-aided measurement, that are very helpful for nearly all beginning researchers and can be taught within the laboratory classroom.  It is beneficial for students and efficient for research advisors to have students develop these abilities as part of their coursework so that research mentors may focus their one-on-one efforts on the unique challenges of a research project.

Regarding the particular choice of laboratory activities, the faculty consensus at CU was that our existing suite of laboratories was adequate.  The faculty felt it would be impossible to span the large range of topics that students may encounter in future research labs.  Therefore, any lab that fits our available resources and expertise, is relevant for some subfield of physics, and enables us to meet our broad learning goals (i.e., modeling, design, communication, and technical lab skills) is a good choice.  It doesn't matter whether or not the particular experiment is close to the cutting edge of physics.  As it happens, the polarization of light lab is rich in opportunities for modeling (we have students use the Jones matrix formalism\cite{Jones1941}) and experimental design and the topic is essential for any student working in optics or using optical detection techniques.   So despite having 200 years of history, the polarization of light lab is still an excellent choice for our advanced lab course. Regarding the choice of apparatus and equipment, for similar reasons we focused on measurement tools and techniques that are relevant for some subfield of physics and enable us to meet our other learning goals.   For the polarization of light lab, some of our equipment was sufficient (standard commercial optomechanics, optical components, HeNe lasers, amplified photodetectors), but other equipment was not adequate.  For instance, one learning goal from the technical lab skills category was that students should be able to perform common measurements using oscilloscopes and acquire data using LabVIEW.  We could not meet this learning goal with our current equipment, so we invested in a larger set of general purpose measurement and data acquisition equipment (oscilloscopes, waveform generators, multimeters, and USB-interfaced data acquisition devices) so that all students would have frequent access to the tools.  

The final question about the appropriate length of the lab activity is partially addressed by our faculty consensus that there was no essential set of required labs.  Since there is flexibility in the choice of labs, the number of labs can be chosen to create experiences that best meet our learning goals.  Currently, we have five guided labs, each two weeks long, followed by a five week final project driven by students' interests.  Students spend five hours in class per week with the instructor and an additional 5-10 hours each week working outside of scheduled hours, some in lab, some elsewhere.  We are currently evaluating our course to determine whether two weeks provides sufficient time to meet our goals or whether students would benefit from longer experiences on the same topic.

The implications of this overall philosophy are liberating for the instructor.  Our instructors are free to choose from a wide range of topics and apparatus in order to create labs that engage students in the practices of scientists.  The main criteria are that the apparatus and physics ideas should be relevant to students as they transition into the community of research physicists (including academia and industry).  

\begin{table*}[t]
    \caption{A comparison of our philosophy of lab instruction and how it compares to a good undergraduate research experience.  Attributes of good research experiences are based on the literature on UREs.\cite{Thiry2011}}
    \begin{tabular}{|p{0.16\textwidth}| p{0.4\textwidth} |  p{0.4\textwidth} |}
        \hline
	& \textbf{A good research experience} & \textbf{A good lab course} \\ \hline\hline
	\textbf{Main goal} & Answering the research question. & Engaging in scientific practices that can be applied later in research. \\ \hline
	\textbf{Problem \newline relevance} & Authentic problems of interest to the broader community.  New results anticipated. & Results will not be new to the broader community, but should be important to know as a member of that community. \\ \hline
	\textbf{Problem \newline complexity} & Open-ended problems, multiple solutions. & Problems have well-defined scope, but designed with some freedom in mind.\\ \hline
	\textbf{Environment} & Research laboratory. & Lab classroom. \\ \hline
	\textbf{Apparatus and  content} & Sophisticated enough to do original research. & Generally useful for research, but often more limited. \\ \hline
	\textbf{Mentoring} & 1-1 mentoring relationships.  An apprenticeship with a master scientist. & Many-to-one teaching relationship with an expert experimentalist. \\ \hline
	\textbf{Community} & Professional physicists (graduate students, post-docs, faculty). & Primarily other students in the class.\\ \hline
          \end{tabular}
    \label{tab:URE_vs_Lab_course}
\end{table*}

\section{A framework for modeling in labs}
\label{sec:Modeling}

In discussions with faculty, modeling emerged as a broad goal that united conceptual and quantitative reasoning within the laboratory course.  Once the general idea was formulated, it needed to be articulated in sufficient detail that it could be integrated into the classroom and assessed.  A common refrain from faculty in support of modeling was that students often complete labs without thinking about the physics. Every laboratory instructor wants his or her students to engage deeply with the physics.  For the junior- and senior-level physics majors in the advanced lab, we wanted this deep engagement to include a significant quantitative component in addition to the underlying conceptual ideas.  Many of our aspirations for promoting deep understanding are already well-articulated in the ideas of models and modeling, which have gained a significant following both within physics education (e.g. Modeling Instruction,\cite{Wells1995} ISLE\cite{Etkina2007a}), more broadly in science education (e.g., the Next Generation Science Standards\cite{Achieve2013}), and even outside the natural sciences (e.g., economic modeling).  The ideas of models and modeling are useful in such a wide variety of circumstances because models serve as a connection between our thoughts (cognitive constructs) and the complexities of the real world.  Models simplify our view of the real world by restricting our attention to the most essential aspects of some real-world phenomenon.\cite{Etkina2006}  And, the various explicit representations of these models, such as diagrams, words, equations, and graphs, give us a way to externally articulate and communicate our thoughts and ideas.  Models are simplified compared to the real-world, but have an added degree of permanence and tangibility compared to our thoughts.  This combination of simplicity and tangibility leads to their use as powerful experimental tools for creating scientific explanations, communicating those ideas, and refining understanding over time.   Modeling is a broad set of activities involving constructing models, using them to make predictions and explanations, using them to interpret data, comparing predictions and data, and refining models based on new evidence.\cite{Etkina2006,Schwarz2009}

The rich history of models and modeling in physics is exemplified by the following quote from Nobel Laureate Willis Lamb.  When reflecting back on his introductory laboratory course at the University of California at Berkeley in 1930, he said, ``I decided that the most important object of physics was to study interesting laboratory phenomena, and to try to make a mathematical model in which the mathematical symbols imitated, in a way to be determined, the motions of the physics system.''\cite{Lamb2001}  Within physics education, modeling came to the forefront through the work of David Hestenes and his collaborators who founded the Modeling Instruction curriculum.\cite{Hestenes1987,Wells1995,Halloun2004,Brewe2008}  Hestenes believed modeling was at the center of scientific activity: ``The great game of science is modeling the real world, and each scientific theory lays down a system of rules for playing the game.''\cite{Hestenes1992}  Since then, modeling has become an explicit part of other curricula, such as the Investigative Science Learning Environment,\cite{Etkina2007a} RealTime Physics,\cite{Thornton1990} and computational modeling.\cite{Buffler2008}

\begin{figure*}[ht]
\centering
\includegraphics[width=0.7\textwidth, clip, trim=0mm 0mm 0mm 5mm]{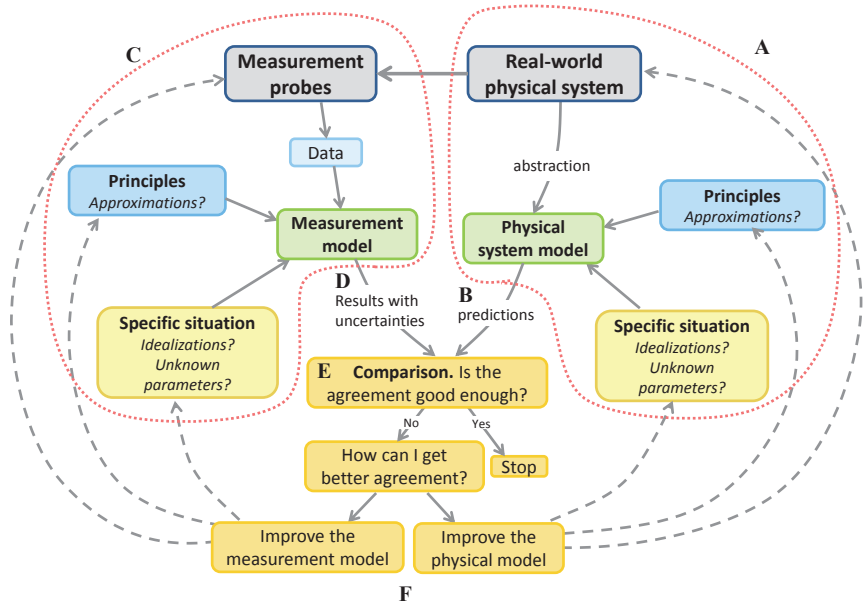}
\caption{A framework for modeling in a physics laboratory involving (A) construction of a model of a real-world physical process, (B) making predictions about the behavior of the physical system, (C) creating a model of measurement tools, (D) using the measurement model to interpret the data and understand the limitations of the measurements, (E) make comparisons between the data and predictions, and (F) model refinement.}
\label{fig:Modeling_framework}
\end{figure*}

In this work, we expand on those ideas that have proved so fruitful in introductory-level physics and create a framework for modeling that can be naturally applied at the upper-division level.  The framework is an organizational tool that helps describe the process of modeling as it relates to doing experiments and describes how the modeling might occur in a lab activity.   There are two primary ways in which this new framework extends earlier discussions of modeling in the literature.  First, it emphasizes both modeling the physical process of interest \textit{and} the measurement tools needed to probe it.  Understanding the measurement tools is often minimized as a learning goal at the introductory level.  However, in the upper-division, as preparation for research in physics, it is essential for students to have an understanding of the principles of operation and limitations of the measurement tools.  This understanding of measurement tools is a helpful prerequisite for engaging in experimental design.  The second difference is that the primary goal of modeling in upper-division lecture and lab courses is usually not to uncover new basic principles.  While an introductory mechanics course may develop the core principles of mechanics (e.g., Newton's laws) through a process of guided inquiry, such an approach is much less common at the higher levels.  Instead, models are constructed by applying known principles (e.g., Maxwell's equations) to describe a specific, possibly complex, real-world phenomenon (e.g., radiation of a oscillating electric dipole moment).  This use of models prioritizes the mapping of the real world onto the abstract model, making simplifications explicit, making predictions, determining the impact of the simplifications on the design of the experiments, identifying and quantitatively modeling systematic error sources, and more.   Figure\ \ref{fig:Modeling_framework} provides a graphical view of the modeling process for laboratory investigations.  

An added benefit of the emphasis on modeling is that it provides a natural way to discuss systematic error in the laboratory.   The development of a well-articulated model should include explicitly identifying the assumptions, idealizations, and approximations that are made.  Systematic error arises when one of these assumptions or approximations is not valid for the phenomenon being investigated.  Although one of the most widely used textbooks on error analysis gives the impression that students are able to do little more suggest a list of possible sources of error,\cite{Taylor1997} through the explicit use of models, students are able to connect these suggestions to the assumptions and idealizations in the model, evaluate the appropriateness of any assumptions or idealizations, and refine the model or apparatus to reduce the systematic error.
 
\section{Developing lab-specific learning goals using the modeling framework}
\label{sec:lab_specific_goals}

In this section, the modeling framework in Fig.\ref{fig:Modeling_framework} is described in detail, and at each stage of the modeling process, examples are given for how it was used to develop lab-specific learning goals for the polarization of light lab.  

Although there are multiple approaches for modeling polarized light, we chose to emphasize the Jones formalism because it can be applied to the simplest experiments involving a single polarizer and for more sophisticated experiments involving multiple polarizers and wave plates at variable angles.  Such experiments are quite easy to set up in the lab, and the Jones formalism provides students with the corresponding theoretical tools to develop predictive models.  For instructors who want to make additional connections across the curriculum, the Jones formalism bridges nicely to modeling two-state quantum systems.\cite{Beck2012}

\begin{figure*}[ht]
\centering
\includegraphics[width=0.8\textwidth, clip, trim=0mm 0mm 0mm 0mm]{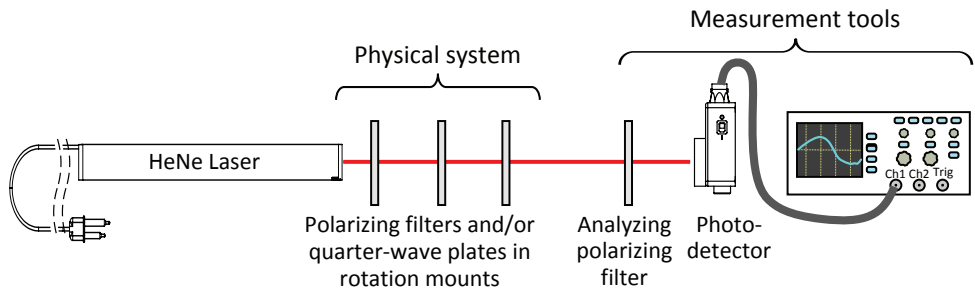}
\caption{The real world components of the polarization of light lab, which need to be understood via modeling.  Laser and detector schematics used with permission from Thorlabs Inc.}
\label{fig:Polarization_Lab_Setup}
\end{figure*}

The first aspect of modeling is \emph{construction}  (sometimes called ``building'' or ``developing'' models).  Construction seeks to answer the following questions: \textit{What real-world phenomena do I want to understand?  What aspects of the real-world system will I include in the model?  What will I ignore?  What principles are needed to describe the phenomena? What approximations or idealizations will I make?  And what parameters are needed in the model?}  Model construction needs to be considered both for the ``physical system'' (Fig.\ \ref{fig:Modeling_framework}A), which is the beam of light and the various optical components that it passes through shown in the left of Fig.\ \ref{fig:Polarization_Lab_Setup}, and the ``measurement tools''(Fig.\ \ref{fig:Modeling_framework}C), which includes an analyzing polarizer on a rotation mount followed by a commercial amplified photodetector connected to a digital multimeter or oscilloscope shown in the right of Fig.\ \ref{fig:Polarization_Lab_Setup}.  

Some of the lab-specific learning goals for model construction focus on the physical system (the state of light) and how it is described using Jones formalism.\cite{Jones1941}  Students should be able to articulate what is included in the model---the polarization state of light---and what is ignored---the spatial distribution of the beam, the oscillations of the electric field as a function of time, etc.  To model the wave plate, students should be able to articulate the principles needed to describe the propagation of light through a dielectric medium (vector wave equation in a linear dielectric medium).  For the polarizer and quarter wave plate, students should be able to relate the macroscopic optical properties to the structure and composition of the material (What is it made of?  What symmetry does the material possess?).  Similarly, students should be able to relate the primary parameters in the Jones formalism to physical quantities in their setup, such the amplitude and phase of the $x$- and $y$-components of the electric fields.  Students should be able to state the idealizations in the model and discuss whether that idealization is valid in their particular experimental setup.  For example, students should recognize idealizations such as the light source is monochromatic, the polarization of the entire wavefront is uniform, and the optical properties of the polarizers and quarter-wave plates are assumed to be uniform over the entire optical element.   Students should also be able to articulate limitations of the model.  For example, the Jones formalism breaks down when combining beams of different wavelengths or spatial profiles.  

Although this emphasis on modeling substantially increases the amount of theoretical detail in the lab, it is all theoretical detail that is intimately related to the design, construction, execution, and interpretation of the experiment.  We want to be clear that a detailed process of building up models from first principles is not always necessary, but it is important that the student treat the model as more than just an equation.  A single physics equation has an astounding amount of physical meaning embedded in a compact symbolic form.  Modeling gives us a process to make the physical meaning in the equation more prominent.   

The next lab-specific learning goal is being able to make accurate quantitative \emph{predictions} of observable phenomena (Fig.\ \ref{fig:Modeling_framework}B).  This means applying the Jones formalism in a variety of situations to model the change in the polarization state of light as it passes through a series of wave plates and polarizers on the way to a detector (Fig.\ \ref{fig:Polarization_Lab_Setup}).  This also means being able to link aspects of the actual experiment to the abstract model (e.g., how does the order of optical elements in the beam path relate to the order of matrix multiplication in the Jones formalism?)  These predictions can be made analytically or, for more sophisticated optical systems, using computational mathematics software (e.g., Mathematica or MATLAB).  We have students begin by using the formalism to predict something familiar, like Malus' law, before moving on to more sophisticated examples involving wave plates.

In addition to the construction of predictive models of the polarization state of light, another lab-specific goal is that students should construct models of the measurement tools (Fig.\ \ref{fig:Modeling_framework}C).  Students should be able to model how light incident upon the photodetector (an oscillating EM field with amplitudes described by a Jones vector) is converted to a measured voltage output from the photodetector.  When measuring polarization states, an analyzing polarization filter is placed in front of the photodetector (as in Fig.\ \ref{fig:Polarization_Lab_Setup}).  Students should be able to model how measurements of voltage versus analyzer angle can be used to determine the elliptical polarization state of the incident light.  This detailed model of the measurement tools is used to interpret the raw voltage vs angle data (Fig.\ \ref{fig:Modeling_framework}D) so that the observations can be directly compared with the predictions.  Just as we want to move students from thinking about equations to thinking about models, we want them to move from treating measurement tools as ``black boxes'' to understanding those tools and their limitations through models. It should be noted that models of the measurement tools are often informed by documentation from the manufacturers, which can include principles of operation and key model parameters and limitations on the ideal functioning of the apparatus.  Thus, another lab-specific learning goal is that students should be able to interpret the manufacturer's documentation for the photodetector and use it to appropriately design and conduct their experiments. 

After predictions and data have been put a similar form, a \emph{comparison} can be made (Fig.\ \ref{fig:Modeling_framework}E).  The comparison of measurement and theory has long been considered the focal point in the laboratory experience, but the modeling framework now places comparison within a larger activity of laboratory sense-making.  The first step in the comparison is deciding the appropriate representation for comparing measurements and predictions.  For instance, students should be able to decide whether the comparison is better represented as a plot of photodetector voltage vs angle, or alternatively as a comparison between the parameters of the elliptical polarization state.  Students should be able to justify why a particular representation is suitable for their goal, especially considering which representation will most effectively communicate their results to colleagues.\cite{Kohl2008}  Students should also be able to use the traditionally emphasized analysis tools, such as estimating uncertainties in measured quantities and propagating uncertainties for derived quantities, but these are now a means to an end---letting students determine whether any deviation between the measurements and predictions are random or systematic.  In this comparison phase, students should be able to analyze and fit models to data using standard software (e.g., Mathematica, MATLAB, or Origin).  An example of a lab-specific learning goal for fitting is that when fitting data of measured power vs analyzer angle, students must be able to justify how many parameters are included in their fit and explain the physical significance of each parameter.  

The final aspect of the modeling framework is iterative model \emph{refinement} (Fig.~\ref{fig:Modeling_framework}F).  After students have made a comparison between predictions and measurements, systematic errors often emerge.  It is natural to consider the limitations and idealizations that were included in the model.  Students should be able to use the idealizations and approximations identified in the model construction phase as starting point for improving the experiment and reducing the systematic error.  As the framework suggests, possible improvements include modifications to the apparatus \emph{and} the models.  Because our experiments using linear polarizers and quarter-wave plates have observable deviations from the idealized Jones matrix model, students should be able to refine the idealized models of linear polarizers and quarter-wave plates, and determine if those refined models better explain their own measurements.  

\section{Implementing Modeling Learning Goals in the Polarization of Light Lab}
\label{sec:Engaging}

While these lab-specific goals describe what we want students to be able to do, they need to be integrated into a coherent lab activity.  In order to engage students in modeling in the laboratory, there are some helpful prior conditions that should be met.  It is important that the sophistication of the physical models and the measurement tools should be appropriate given the amount of time students have available for doing the experiment and given their prior knowledge of the physics ideas and apparatus.   If the content of the lab is too sophisticated, the lab becomes more about following instructions, either from the lab guide or instructor, rather than an effort of deep sense-making.  The polarization of light lab is appropriate because the ideas of polarization and waves are familiar from lecture courses in E\&M and an elective lecture course on optics.  Similarly, the equipment involves basic optical components, optomechanics, and detectors that students are already using for a variety of other optics labs throughout the semester.

We want to build activities that naturally engage students with sophisticated modeling practices described in Secs.~\ref{sec:Modeling}~and~\ref{sec:lab_specific_goals}.  Our current approach is to use a form of guided inquiry.\cite{Fuller1977,Bruck2008}  There are many reasons to suppose that some form of explicit prompting is a good solution.  First, there is existing research from cognitive science\cite{Collins1991,Mayer2004,Kirschner2006} and from attempts to scaffold the learning of scientific practices\cite{Hmelo-Silver2007,Etkina2008} that indicates such guidance is an efficient way to develop expertise. Second, a similar approach is taken by other tested physics curricula, such as Physics in Everyday Thinking,\cite{Jenness2008} the Investigative Science Learning Environment,\cite{Etkina2008} and Modeling Instruction.\cite{Halloun1987}  Third, we have faculty with a wide range of backgrounds teaching our lab course and we need course materials that support all instructors in helping students develop modeling abilities.   Fourth, we have evidence from classroom observations that questions with overly vague directions and goals can leave students with unclear expectations and set a very low bar for achievement, although this can be overcome through intensive and individualized instructor-student interactions.  The things that differentiate our guided inquiry from the often-maligned ``cookbook labs'' is that cookbook labs require very simple procedural responses to the prompts, similar to questions that elicit low-level cognition as categorized by Bloom's Taxonomy.\cite{Domin1999a,Krathwohl2002}  However, the prompts in our labs aim for larger steps in reasoning and higher-levels of thinking than is common in cookbook labs.  For each of the various aspects of modeling outlined in Sec.\ \ref{sec:lab_specific_goals}, we present example questions or prompts that are used in our polarization of light lab.  

The prompts shown below have been used as part of a longer laboratory activity that extends over two weeks.  Over those two weeks, students build the experiments, develop predictive models that are implemented computationally, and test those models against their data.  We have run this lab activity during three different semesters of the CU advanced lab with students in their junior year or beyond.  

\textbf{1. Construction} prompts for the physical system (Fig.\ \ref{fig:Modeling_framework}A) involve students in identifying the key elements of the system, what is being ignored, the physical principles, and other parameters that connect the model to the specific details of the experiment.   

\textit{\textbf{Example:}  What basic physics ideas explain why the polarizing filter only absorbs one polarization?  What makes a quarter-wave plate different than glass?} It is \emph{possible} to take data and fit it to an equation without understanding how the polarizing filter works, but an experimentalist loses credibility when s/he cannot explain how key elements of the apparatus function. 

\textit{\textbf{Example:}  Suppose two beams of light of different polarization $\left(\begin{matrix} E_x \\ E_y \end{matrix}\right)$ and $\left(\begin{matrix} E^\prime_x \\ E^\prime_y \end{matrix}\right)$ are combined using a beam splitter.  The Jones matrix model suggests that the final polarization state after a 50/50 beamsplitter would be proportional to $\left(\begin{matrix} E_x + E^\prime_x \\E_y +  E^\prime_y \end{matrix}\right)$. Under what experimental conditions would this use of the Jones formalism be valid, meaning it would accurately describe the final polarization state of the combined beams of light after the beamsplitter?}  This question targets the simplifications and limitations in the model.

\textbf{2. Prediction} prompts (Fig.\ \ref{fig:Modeling_framework}B) encourage students to make quantitative predictions that can be compared to measurements or a previously known result. In this example, the prediction should be familiar from introductory physics---Malus' Law for crossed-polarizers.

 \textit{\textbf{Example:}  Using the Jones formalism, derive Malus' law (the transmitted power through two crossed polarizers).  This exercise will give you confidence in applying the Jones formalism to more complicated models, like those using the quarter-wave plate.}  Note that ``derive'' in this case is really just another form of model prediction.

\textbf{3. Constructing models of the measurement tools} and using them to \textbf{interpret data} (Fig.\ \ref{fig:Modeling_framework}C \& D) encourages a deeper understanding of the apparatus.  In the polarization lab, a model is needed to connect the raw data of photodetector voltage as a function of analyzing polarizer angle into an estimate of the parameters of elliptical polarization state.  

\textit{\textbf{Example:}  Predict the power transmitted through the analyzing filter as a function of the filter's orientation for an arbitrary incident elliptical polarization state.  Use the predictive function as a fit function for your real data.} 

\textbf{4. Comparison} between measurements and predictions (Fig.\ \ref{fig:Modeling_framework}E) can be made at a number of points during the activity, but here is one example related to creating circularly polarized light.

\textit{\textbf{Example:}  Predict the elliptical polarization parameters for linearly polarized light passing through a quarter wave plate.  What angle between the wave plate axis and the linear polarization produces circularly polarized light?  Experimentally attempt to produce circularly polarized light.  What are the measured elliptical polarization parameters for the light that is intended to be circularly polarized?}

\textbf{5. Refinement} of the models, measurements, and apparatus (Fig.\ \ref{fig:Modeling_framework}F) can occur in many ways, but for our particular experiment we knew from previous experience that students typically produce near-circularly polarized light that has easily observable systematic deviations.   In this prompt, students are asked to extend their existing model to include non-ideal features that could account for the deviation and use them to make new predictions.  The goal is to see if students can distinguish between the effects of possible systematic error sources and quantitatively model the systematic error.  The ability to treat systematic error as a natural part of the lab course and as a source of open-ended inquiry are two virtues of our modeling framework.  A statement of possible systematic error sources is no longer the end of the laboratory activity,\cite{Taylor1997} but is the beginning of a rich inquiry into refinement of the models and apparatus.

\textit{\textbf{Example:}  This question explores the systematic error effects that could limit your ability to produce circularly polarized light.  (a) Predict how a small violation of the idealization would change the result. (b) Can you distinguish between the systematic error sources? (c) Could this systematic error account for your non-ideal result?  (d) Is the violation of the idealization within tolerances on our ability to measure angles or the specifications on the quarter-wave plate?  (e) Which error source, if any, is most likely?} 

These prompts are all in the context of a lab on the polarization of light, but we hope they give a sufficient impression of the breadth of modeling-related questions that could be incorporated into any laboratory activity.

Students who have worked through these prompts and master the apparatus, data collection, and modeling have found the polarization of light lab quite satisfying.  Some groups were challenged by the mathematical and computational modeling.  Part of the challenge is integral to the lab experience---connecting the mathematical formalism to the real world apparatus and measurements.  Additionally, some groups have found the computational modeling challenging, but we hope this will be alleviated in future semesters as more students are gaining experience with computational techniques earlier in the curriculum in courses such as classical mechanics. 

\section{Conclusions}

In the beginning of the paper, we posed a number of questions that arise when thinking about laboratory course transformation.  We gradually worked through these questions by starting with a teaching philosophy and broad course goals and successively refined these goals until we arrived at individual prompts in a particular lab guide.  The modeling framework (Fig.\ \ref{fig:Modeling_framework}) graphically portrays a generally-applicable approach for incorporating modeling into the learning goals and activities of a laboratory course.  Although learning goals beyond those encompassed by modeling are also relevant, we chose to focus this paper on modeling because it forms a valuable framework for emphasizing students' sense-making in the laboratory.  

In parallel to using the modeling framework for the construction of curricula, we are also using it as a guide for interpreting students in-the-moment thinking in the laboratory.  In a preliminary analysis of individual students doing a 30 minute lab activity, we have already seen the benefits of the dual emphasis on the measurement tools and the physical system.  In this short activity, students spent about twice as much time focused on the measurement tools as compared to the physical system when trying to compare their data with predictions.  Their focus on the measurement tools included identifying the principles of operation, using the tools to make measurements, and using a model of the measurement tools to interpret data.  Also, we found that students often did not explicitly articulate the simplifications, assumptions, and approximations in their models during the construction phase.  Students who did not articulate these simplifications had difficulty making connections to limitations of their measurements and predictions that may need to be reevaluated in future iterations of the experiment.  An implication is that our guided curricula may need to more frequently remind students to articulate simplifications, approximations, and assumptions as they construct models.

Lastly, our approach and philosophy raise many interesting research questions about pedagogy in laboratory courses.  We are conducting ongoing research into how course artifacts like lab notebooks, written reports, and oral presentations, can be used to assess learning goals.  As we come to better understand these standard laboratory assessments, we can examine how different kinds of guidance within the lab support different levels of achievement of our learning goals, and especially importantly, whether students are able to transfer their demonstrated abilities in a classroom laboratory into the research laboratory.

\section{Acknowledgments}
The authors would like to thank the CU Physics Department for contributing to the learning goals process.  We would also like to particularly thank faculty members Debbie Jin, Charles Rogers, Noel Clark, and Kevin Stenson for contributing their teaching insight and experimental expertise as they gave feedback on the previous and newly revised advanced lab course.  This work is supported by NSF CAREER PHY-0748742, NSF TUES DUE-1043028, JILA PFC PHY-0551010, the CU Science Education Initiative, and Integrating STEM Education at Colorado NSF DRL-0833364.  The views expressed in this paper do not necessarily reflect those of the National Science Foundation.

\end{document}